\input{aipcheck}

\documentclass[
    ,final            
  ]
  {aipproc}

\layoutstyle{8x11double}

\def\XX{X(3872)}

\def\ddsn{D^{0*}\overline{D^0}} 	
\def\ddsc{D^{\pm*}D^{\mp}}		

\def\etab{\end{tabular}}  
  
\def\bit{\begin{itemize}}
\def\eit{\end{itemize}}

\def\bml{\begin{multline}}
\def\eml{\end{multline}}
\def\be{\begin{equation*}} 
\def\ee{\end{equation*}}  
\def\bea{\begin{eqnarray*}}  
\def\eea{\end{eqnarray*}} 
\def\bmu{\begin{multline}}
\def\emu{\end{multline}}
\def\bal{\begin{array}{l}} 	
\def\eal{\end{array}}

\def\MeV{\textrm{ MeV}}

\def\letterS{S}
\def\letterP{P}
\def\letterD{D}
\def\letterF{F}
\def\greek#1{\def\letter{#1}
\ifx\letter\letterS\Sigma
\else
	\ifx\letter\letterP\Pi
	\else
		\ifx\letter\letterD\Delta
		\else
			\ifx\letter\letterF\Phi
			\else XXX
			\fi
		\fi
	\fi
\fi
}

\def\uu{u\overline u}

\def\dd{d\overline d}

\def\dd{d\overline d}

\def\cc{c\overline c}

\def\cc{c\overline c}

\newcommand{\ket}[1]{|           {#1}           \rangle}
\newcommand{\bra}[1]{\langle           {#1}           |}

\renewcommand{\u}[1]{\textrm{#1}}
\def\cn#1#2#3{^{#1}\u{#2}_{#3}}     
\def\an{\cn} 

\def\Jp{J/\psi}

\def\xQN#1,{\def\Element{#1}%
\ifx\Element\endpiece
\else 
	\if\Element/\def\ApplyOrLiteral{1} 
	\else
		\if\ApplyOrLiteral0\ElementRule\Element 
		\else\Element\def\ApplyOrLiteral{0}
		\fi
	\fi
	\expandafter\xQN
\fi}

\def\xNQ#1,{\def\ElementRule{#1}%
\ifx\ElementRule\endpiece
\else 
	\if\ElementRule/\def\ApplyOrLiteral{1} 
	\else
		\if\ApplyOrLiteral0\ElementRule\Element 
		\else\ElementRule\def\ApplyOrLiteral{0}
		\fi
	\fi
	\expandafter\xNQ
\fi}

\def\k|#1>{\ket{#1}} 
\def\b<#1|{\bra{#1}} 

\def\bk<#1|#2>{\langle #1 | #2 \rangle}
\def\Bk#1<#2|#3>{\def\ElementRule{#1}\def\ApplyOrLiteral{0}
\xQN/,\langle,#2,xxx,\vert #3 \rangle}
\def\bK#1<#2|#3>{Undefined}

\def\BK#1<#2|#3>{\def\ElementRule{#1}\def\ApplyOrLiteral{0}
\xQN/,\langle,#2,/,|,#3,/,\rangle,xxx,}
\def\KB<#1|#2>#3{\def\Element{#3}\def\ApplyOrLiteral{0}
\xNQ/,\langle,#1,/,|,#2,/,\rangle,xxx,}

\def\me<#1|#2|#3>{\langle #1\vert #2\vert #3\rangle}
\def\ME#1<#2|#3|#4>{\def\ElementRule{#1}\def\ApplyOrLiteral{0}
\xQN/,\langle,#2,/,\vert,/,#3,/,\vert,#4,/,\rangle,xxx,}
\def\EM<#1|#2|#3>#4{\def\Element{#4}\def\ApplyOrLiteral{0}
\xNQ/,\langle,#1,/,\vert,/,#2,/,\vert,#3,/,\rangle,xxx,}

\def\rme<#1|#2|#3>{\langle #1\Vert #2\Vert #3\rangle}
\def\RME#1<#2|#3|#4>{\def\ElementRule{#1}\def\ApplyOrLiteral{0}
\xQN/,\langle,#2,/,\Vert,/,#3,/,\Vert,#4,/,\rangle,xxx,}
\def\EMR<#1|#2|#3>#4{\def\Element{#4}\def\ApplyOrLiteral{0}
\xNQ/,\langle,#1,/,\Vert,/,#2,/,\Vert,#3,/,\rangle,xxx,}

\def\endpiece{xxx}

\def\be{\begin{equation}}
\def\ee{\end{equation}}
\def\bea{\begin{eqnarray}}
\def\eea{\end{eqnarray}}

\def\XX{X(3872)}
\def\ddsn{D^{0*}\bar{D}^0} 
\def\ddsc{D^{\pm*}D^{\mp}} 
\def\an#1#2#3{^{#1}\textrm{#2}_{#3}}     

\def\k|#1>{\ket{#1}}
\def\Jp{J/\psi}
\def\b<#1|{\bra{#1}} 
\def\bk<#1|#2>{\langle #1 | #2 \rangle}

\begin{document}

\title{Rethinking the X(3872)}


\author{T. J. Burns \footnote{\tt Timothy.Burns@roma1.infn.it}}{
  address={INFN Roma, Piazzale A. Moro 2, Roma, I-00185, Italy}
}
\begin{abstract}
The BaBar analysis which favours $2^{-+}$ quantum numbers for the $X(3872)$ implies that it may be none other than the $1^1$D$_2$ charmonia state. In that case the isospin breaking in closed flavour modes may be the result of re-scattering from open flavour. However, the observed production cross section in proton-antiproton collisions is much larger than expectations, while the mass of the state, compared to the predictions of a string model, is too high. The $\an 1D2$ assignment would imply a $\an 3D2$ partner nearby in mass, which may be the $X(3875)$. In the tetraquark interpretation the $2^{-+}$ assignment implies a rich spectrum of partner states, although the $\XX$ may be among the few which are narrow enough to be observable. This talk is based in part on ref. \cite{BurnsPiccininiEtAl102-+}. 
\begin{center}
Contribution to the proceedings of {\it Quark Confinement and the Hadron Spectrum IX},\\
Universidad Complutense de Madrid, 30 August--3 September 2010.
\end{center}

\end{abstract}

\maketitle


\section{Introduction}

In a recent paper the BaBar collaboration claims~\cite{Amo10evidence} that the quantum numbers of the $X(3872)$ are not $1^{++}$, as had generally been accepted, but $2^{-+}$. Although a high statistics analysis by CDF \cite{AbulenciaEtAl07analysis} allowed both possibilities, an earlier conference paper by Belle~\cite{AbeEtAl05experimental} favored $1^{++}$, and most theoretical work has been based on this assumption. The new result from BaBar therefore provokes a revision of theoretical ideas concerning the $\XX$. If this quantum number assignment is confirmed, the molecular interpretation of the $X(3872)$ as a loosely bound state of a $D^0$ and a $\bar D^{*0}$ meson seems unlikely. Whereas a $1^{++}$ state can be formed  in relative S-wave, a $2^{-+}$ state would require a relative P-wave, and it is unlikely that $\pi$ exchange could bind such a state, given that even in S-wave it is not clear that the attraction is sufficiently strong~\cite{Suzuki05x(3872)}. 

\section{The $\cn 1D2$ interpretation}
With the molecular interpretation rendered less likely, the BaBar result revives the possibility that the $\XX$ is a conventional $1\an 1D2$ charmonium state. The main challenge is the strong violation of isospin: the $X(3872)$ decays to $J/\psi\rho$ and $J/\psi\omega$ with approximately equal magnitude \cite{AbeEtAl05evidence}. It was noted long ago \cite{ClosePage04d*0} that due to its proximity to the threshold, the $\XX$ ought to couple more strongly to $\ddsn + c.c.$ than to $\ddsc$, which lies some 8 MeV higher. In the molecular interpretation this leads to hidden light quark content $\uu$, which is a state of mixed isospin:
\be
\ket{\cc\uu}	
=\frac{1}{\sqrt 2}\ket{\cc\frac{\uu+\dd}{\sqrt 2}}+\frac{1}{\sqrt 2}\ket{\cc\frac{\uu-\dd}{\sqrt 2}}.
\label{isospin}
\ee
In the limit that the $\dd$ component is completely suppressed and that the $\rho$ and $\omega$ have equal mass and width, one thus predicts equal magnitude isospin one and isospin zero $\Jp\rho$ and $\Jp\omega$ modes. Even if the $\XX$ is a standard $\cc$ meson, however, an essentially very similar argument may explain the observed isospin breaking. 

Suppose that the $\XX$ lies slightly above the $\ddsn$ threshold. The strong decay $\XX\to \ddsn +c.c.$ is allowed, while $\XX\to \ddsc$ is forbidden; this is a maximal violation of isospin, which would require neutral and charged modes in equal measure. The produced $\ddsn+c.c.$ could then re-scatter into a charmonia plus light meson, and the flavour content of the latter would thus be an equal admixture of isospin zero and isospin one. Moreover, since their thresholds are remarkably similar to the $\XX$ mass, at 3872.41 MeV and 3879.57 MeV respectively, the $\Jp\rho$ and $\Jp\omega$ pairs have small momenta, and thus the re-scattering would require very little momentum transfer from the small momentum $\ddsn+c.c.$ pair. A similar mechanism could be in place if the $\XX$ is slightly below threshold, in which case the dominance of the neutral charmed mesons (and hence the maximal isospin violation) is due to the large energy denominator; this is essentially the idea proposed in ref. \cite{Suzuki05x(3872)} in the context of the $1^{++}$ $\XX$.


As has been widely discussed in the literature \, potential models typically predict a $1\an 1D2$ state lying some 50-100 MeV lighter than the $\XX$ \cite{BarnesGodfrey04charmonium,EichtenLaneEtAl04charmonium,EichtenLaneEtAl06new}. In a string model one obtains a similar prediction; in the heavy quark limit one finds an expression for the mass $E$ of a meson in terms of its orbital angular momentum $L$,
\be
E= 2m\left(1+\frac{3}{2}\left(\frac{LT}{2m^2}\right)^{2/3}\right),
\ee
where $m$ is the quark mass and $T$ the string tension. The corresponding result for light quarks yields the standard Regge trajectories $E\sim \sqrt{TL}$ \cite{SelemWilczek06hadron}.

Assuming that the $L=0$ state of the string is described by the spin-weighted average of the $\an 1S0$ and $\an 3S1$ states, there is a linear relation among the masses,
\be
M(\an 1D2)=2^{2/3}M(\an 1P1)+\frac{1-2^{2/3}}{4}\left(M(\an 1S0)+3(\an 3S1)\right)
\ee
and a parameter-independent prediction of the mass of the $\an 1D2$ state in terms of the masses of the $h_c$, $J/\psi$ and $\eta_c$,
\be
M(\an 1D2)=3795 \MeV.
\ee
In agreement with potential model predictions, this is difficult to reconcile with the observed $\XX$ mass. The validity of the mass formula can be tested in the bottomonia spectrum, using as input the experimental values for the $\eta_b$ and $\Upsilon$ masses, and for the $h_b$ the centre of gravity of the $\chi_{b0}$, $\chi_{b1}$ and $\chi_{b2}$; the formula yields a predicted mass for the unobserved $\an 1D2$ state $M(\eta_{b2})=10168.72^{+1.4}_{-1.8}$ MeV, to be compared with the spin-weighted average $M(\eta_{b2})=10165.84\pm 1.8\textrm{ MeV}$ of the recently observed \cite{AmoBABAR10observation}  $\an 3D{1,2,3}$ states of the upsilon sector.


In addition to its discovery mode in B decays, the $\XX$ has been observed in proton-antiproton collisions by both CDF \cite{AcostaEtAl03measurement} and D0 \cite{AbazovEtAl04observation}. Ref \cite{BignaminiGrinsteinEtAl09is} considered the hypothesis that the $\XX$ is a loosely bound molecule of $D^0$ and  $\bar D^{*0}$, and found that the predicted cross section is considerably smaller than that which is observed, consistent with expectations that it is difficult to produce a loosely-bound S-wave molecule in a high energy hadron collision environment. If the $X(3872)$ is a $1 \an 1D2$ standard charmonium, the production cross section is also expected to be small. The fragmentation function, which describes the probability that quarks and gluons hadronise into bound states, can be expressed as a perturbative expansion in the quark velocity $v$. For a $1 \an 1D2$ state the function begins at order $O(v^7)$, and so one expects a smaller cross section than for 1P states, for which the function is $O(v^5)$, which are themselves suppressed with respect to 1S states, $O(v^3)$. 

Using fragmentation functions of Cho and Wise \cite{ChoWise95gluon} and recent gluon distribution functions we find a predicted cross section
\begin{equation}
\sigma (p\bar p\to 1\an 1D2+all)=0.6~{\rm nb},
\label{charmxsect}
\end{equation}
some 50 and 120 times smaller than the estimated experimental cross section. It is difficult, therefore, to reconcile the observed production cross-section of the $\XX$ with the expectations for a $1\an 1D2$ state. Other difficulties associated with the $\an 1D2$ interpretation of the $\XX$ are discussed in refs. \cite{KalashnikovaNefediev10x(3872),JiaSangEtAl10is}.


A feature common to most potential models, as can be seen by the comparison of mass predictions presented in ref \cite{BarnesGodfrey04charmonium}, is that the $\an 1D2$ is accompanied by a $\an 3D2$ partner whose mass is within a couple of MeV. Is it possible that this state has already been observed? Both Belle \cite{GokhrooEtAl06observation} and BaBar \cite{AubertEtAl08study} have observed a second state, the $X(3875)$, with a mass a few MeV higher than, and inconsistent with, that of the $X(3872)$. The interpretation of the experimental data is a subtle issue \cite{AushevEtAl10study} and it is possible that the two states are one and the same \cite{DunwoodieZiegler08simple}. Supposing, however, that the two states are distinct: what is known about the heavier one? Whereas the radiative decays of the lighter $\XX$ fix its charge conjugation to be positive, the heavier $X(3875)$ is observed only in $D^0\bar D^0\pi^0$, a mode allowed for both positive and negative charge conjugation states. This raises the intriguing possibility that the $X(3875)$ is the $2^{--}$ state $\an 3D2$. 

Being above threshold, such a state ought to decay dominantly to $D^0\bar D^0\pi^0$, consistent with observations. Moreover the proposed mechanism for the enhanced closed flavour modes $\Jp\rho$ and $\Jp\omega$ in the $\XX$ need not imply a large $\Jp\eta$ in the $X(3875)$, since the momentum transfer would be large, and this is consistent with the current (weak) upper limit \cite{AubertEtAl04observation}. A critical test of the $\an 3D2$ scenario would be the observation of $X(3875)$ in $\chi_1\gamma$. The current experimental upper limit on this mode is not yet strong enough to exclude the possibility, as can be deduced from the arguments of ref \cite{BarnesGodfrey04charmonium} in the context of the lighter $\XX$.

\section{The tetraquark interpretation}

The $2^{-+}$ assignment also changes the outlook for the tetraquark interpretation of the $\XX$, since it would have a unit of orbital angular momentum. The existence of such a state would imply the existence of a rich spectrum of partner states, formed of combinations of spin 1 (axial vector) and spin 0 (scalar) diquarks coupled to the $L=1$ orbital angular momentum to give various total angular momenta $J$. This leads to one each of $0^{--}$ and $3^{--}$, two each of $2^{--}$, $0^{-+}$, $1^{-+}$ and $2^{-+}$, and four $1^{--}$. There ought presumably also to exist a set of lighter S-wave states, two each of $0^{++}$ and $1^{+-}$, and one each of $1^{++}$ and $2^{++}$. Moreover for each $J^{PC}$ there would be various flavour combinations: two neutral and two charged states formed out of $[cu]$ and $[cd]$ building blocks and the corresponding antidiquarks, and a further four strange states if one considers strange diquarks. 

In the absence of experimental evidence for the many predicted partner states a tetraquark explanation for the $\XX$ is only tenable if one can explain why it, among all the possible configurations, is unique. One possibility is that most of the states are so broad as to be effectively unobservable, while the $\XX$ is among the few which are narrow and thus observable. Because of its unnatural parity the $\XX$ cannot decay to $\eta_c\pi$, $\eta_c\eta$ or $D\bar D$; its observed decays into $J/\psi\rho$, $J/\psi\omega$ and $\ddsn$ are all P-wave with very little phase space, implying a small width in accordance with the experimental data. 

The pattern of allowed decays for the $0^{-+}$ states is very similar to those of the $2^{-+}$ states, and one therefore expects that the $0^{-+}$ states should exist and will be as narrow as the $\XX$. Like the $\XX$, they should decay radiatively into $J/\psi\gamma$, and there is apparently no signal in the data \cite{AubertEtAl09evidence}. The $0^{--}$ and $2^{--}$ states will presumably be broader than the corresponding $0^{-+}$ and $2^{-+}$ states due to the much greater phase space available to $J/\psi\pi$ and $\Jp\eta$ compared to $\Jp\rho$ and $\Jp\omega$; these may be too unstable to be identified. The remaining $1^{--}$, $3^{--}$ and $J^{PC}$-exotic $1^{-+}$ states have many decay modes available and with ample phase space, so we expect that they should be the least stable of all the possible $L=1$ tetraquark configurations. This may be helpful from a phenomenological point of view, since there is no evidence of an overpopulation of states in the well-studied $1^{--}$ sector in the 3800-3900 MeV mass region.

The positive parity states are difficult to accommodate with data, insofar as they are as light or even lighter than the corresponding $h_c$ and $\chi_c$ states, sharing the same quantum numbers and hence decay modes, and yet there is apparently no experimental evidence for their existence. The $0^{++}$ and $1^{+-}$ can decay in S-wave to $\eta_c\pi$ and $J/\psi\pi$ respectively, and one could argue that they simply ``fall-apart'' broadly in such a way as to be effectively unobservable. The remaining $L=0$ states cannot be dismissed in this way. The $1^{++}$ states cannot decay into $\eta_c\pi$ and should be comparatively narrow, while the $2^{++}$ states could decay to $\eta_c\pi$, but only in D-wave and with the non-conservation of heavy quark spin. In the tetraquark picture one thus expects light, narrow $1^{++}$ and $2^{++}$ states in the $\chi_c$ mass region decaying into $\eta_c\pi$ and $J/\psi\gamma$.

\section{Conclusions}
The BaBar result which favours $2^{-+}$ quantum numbers for the $\XX$ implies a serious revision of theoretical interpretations is required. If the $\XX$ is a $1\an 1D2$ state both its mass and production cross section at the Tevatron are larger than expected. In any case, the $\an 3D2$ partner state should exist within a few MeV and may be the $X(3875)$. If the $X(3872)$ is, instead, a P-wave tetraquark, there ought to exist a series of S-wave states with masses comparable to P-wave charmonia, and this is difficult to reconcile with data. 



\bibliographystyle{aipproc}   

\bibliography{tjb}

\IfFileExists{\jobname.bbl}{}
 {\typeout{}
  \typeout{******************************************}
  \typeout{** Please run "bibtex \jobname" to optain}
  \typeout{** the bibliography and then re-run LaTeX}
  \typeout{** twice to fix the references!}
  \typeout{******************************************}
  \typeout{}
 }

\end{document}